\documentclass[aps,prb,superscriptaddress,floatfix, twocolumn,showpacs]{revtex4}
\usepackage{amsmath}
\usepackage{amsfonts}
\usepackage{amssymb}
\usepackage{color}
\usepackage{ dsfont }
\newcommand{\beq}{\begin{equation}}
\newcommand{\eeq}{\end{equation}}
\newcommand{\beqa}{\begin{eqnarray}}
\newcommand{\eeqa}{\end{eqnarray}}
\usepackage{graphicx,amssymb,amsmath,subfigure}
\usepackage{dcolumn}
\usepackage{bm}

\begin{document}
\title{Anomalous Hall Response in Two Dimensional Topological Insulators due to the Stark Effect}

\author{Alberto Cortijo}
\email{alberto.cortijo@csic.es}
\affiliation{Instituto de Ciencia de Materiales de Madrid,
CSIC, Cantoblanco; 28049 Madrid, Spain.}
\date{\today}

\begin{abstract}
It is shown that the presence of matrix dipole moments induced by external electric fields can modify the Hall response in two dimensional Topological insulators. In the case of the Quantum Anomalous Hall effect the induced transverse currents acquire an extra term being proportional to the Hall conductance and the time derivative of the applied electric field. In the case of the Quantum Spin Hall effect both a spin and charge transverse currents appear simultaneously. In virtue of the bulk-edge correspondence the coupling between the chiral edge channels and the electromagnetic field changes allowing for an extra non minimal coupling term. Both effects can be measured through transport and tunnelling experiments.
\end{abstract}

\maketitle
\section{Introduction}

For non interacting systems, the appearance of non dissipative Hall-like currents is due to the existence of topological structures present in the electronic spectrum\cite{TKN82,KM05}. In the case of the quantum Hall effect (QHE) the physical ingredient allowing for such structure is the presence of an external magnetic field breaking time reversal symmetry, while in the case of the quantum spin Hall effect (QSHE) it is the spin-orbit coupling what allows for a spin resolved currents along the edges\cite{SCN04,BZ05}. The quantum anomalous Hall effect (QAHE) lies somewhat in between: time reversal symmetry is broken by the presence of magnetic elements instead and the QSHE can be understood as two copies of the QAHE related by time reversal symmetry\cite{H88}. Remarkably, not only the QHE has been measured\cite{KDP80} but the QSHE and the QAHE have been experimentally confirmed\cite{KWS07,CZF13}. A recurrent question is if this electromagnetically induced response can be modified somehow. It is known that there is no room for such modification unless interactions are present due to the topological meaning of this response\cite{TSG82}. However this is the case for the Hall response in the DC limit. Little is experimentally known when time dependent external electromagnetic fields are considered.Theoretical and experimental efforts have been carried out to understand and measure the frequency structure of closely related responses like the the Faraday and Kerr effects in three dimensional topological insulators (3DTI) and graphene\cite{GJ12,VSL12,SYY13}. Another important property of systems exhibiting Hall responses is the presence of one dimensional conducting channels at the sample's edge. These edge states are the responsible of the conducting properties of these phases of matter when the Fermi level lies in the bulk gap. The transport properties of these conducting channels have been extensively studied in the literature, both for the DC and the AC limits\cite{C03}. Besides its inherent interest in fundamental science\cite{GC11}, it is not necessary to mention the potential applicability of such non dissipative edge channels for future electronic devices both in the DC and in the optical frequency regime, in special for HgTe/CdTe based devices. For these reasons it is interesting to investigate how systems  showing Hall responses (like HgTe/CdTe quantum wells) behave under the effect of time varying electric fields. Also we will see that the presence of external time dependent electric fields can unveil previously unreported properties of such systems, like the one described in the present work when dipole interactions between states close to the Fermi level are considered. Such interactions are the manifestation in Condensed Matter Physics of the Stark effect which is the electrical analogue of the Zeeman effect\cite{V01}.

The rest of the paper is organized as follows: In section II we describe the Bernevig-Hughes-Zhang lattice model including the dipole moment terms and get the continuum version. In section III we calculate the induced Hall current through integrating out fermions and getting the properly modified Chern-Simons term. We also describe here how the chiral edge states in the QAHE and QSHE get modified through the bulk-boundary correspondence. Finally, in section IV we summarize the results obtained.

\section{The model}

\begin{figure} %  figure placement: here, top, bottom, or page
\centering
\includegraphics[width=1.1\columnwidth]{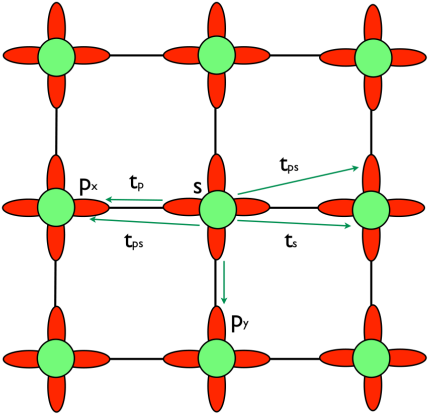}
\caption{(Color online) Scheme of the system described in the text after \cite{NSO10}. The green (light) circles and red (dark) lobules represent $s$ and $p$ orbitals, respectively.}
   \label{fig:lattice1}
\end{figure}

Let us consider as a prototypical model for the QAHE and QSHE the Bernevig-Hughes-Zhang model corresponding to a square lattice with three orbitals per site\cite{BHZ06}. These orbitals are chosen to be s-like ($J=1/2$) and p-like ($J=3/2$). Is then conceivable that when an external electric field is applied, intra-atomic dipole interactions might take place between these orbitals. For convenience we will consider time dependent external electric fields but in the dipole approximation, that is, we do not take into account spatial dependences of the fields: $\mathbf{E}=\mathbf{E}(t)$.
The lattice Hamiltonian in the tight binding approximation in absence of external perturbations reads\cite{BHZ06,FK07,NSO10}:
\begin{eqnarray}
\nonumber H_{t}=&-&\sum_{i,a=x,y}t_{s}s^{+}_{i}s_{i+a}-\sum_{i,a=x,y}t_{p}p^{+}_{a,i}p_{a,i+a}-\\
\nonumber &-&\sum_{i,a=x,y}t_{ps}\left(s^{+}_{i}p_{a,i+a}-s^{+}_{i}p_{a,i-a}\right)+\\
&+&\sum_{i,a=x,y}\epsilon_{s}s^{+}_{i}s_{i}+\epsilon_{p}p^{+}_{a,i}p_{a,i}+h.c.\label{latham}
\end{eqnarray}
To (\ref{latham}) we have to add a spin orbit coupling
\begin{equation}
H_{SO}=\sum_{i}\lambda L^{z}_{i}S^{z}_{i}.\label{SOham}
\end{equation}
The geometry of the hopping terms can be seen in fig(\ref{fig:lattice1}). When the spin-orbit term is considered, the on-site energies change to $E_{0}=\epsilon_{s}$, $E_{\pm1}=\epsilon_{p}\pm\frac{\lambda}{2}$. Assuming $\epsilon_{s}<\epsilon_{p}$ and $\lambda>0$ to be the largest energy scale involved in the problem, we can neglect the $p_{+1}\equiv (p_{x}+i p_{y})/\sqrt{2}$ orbital, drastically reducing the problem to a two band problem. Firstly we will focus on the QAHE so we will work out the model (\ref{latham}) in its angular momentum polarized version\cite{LQD08,YZZ10}.

In order to capture the effect of the electric dipole terms we will consider the standard radiation-matter coupling:
\begin{equation}
H_{d}=-q\mathbf{E}(t)\int d^{2}\mathbf{r}\psi^{+}(\mathbf{r})\mathbf{r}\psi(\mathbf{r}).\label{intham1}
\end{equation}
 In the tight-binding approximation, the fermion operator is written as $\psi(\mathbf{r})=\sum_{i,\alpha}c_{i,\alpha}\phi_{\alpha}(\mathbf{r}-\mathbf{R}_{i})$ ($c_{i,\alpha}$ represent the amplitudes $s_{i}$ and $p_{i,-1}$ and $\phi_{\alpha}$ the orbital wave functions). Inserting this expression in (\ref{intham1}) and assuming that the resulting overlap integrals are nonzero only for the same site, we can write
 
\begin{eqnarray}
\nonumber &&\int d^{2}\mathbf{r}\phi^{*}_{\alpha}(\mathbf{r}-\mathbf{R}_{i})\mathbf{r}\phi^{*}_{\alpha'}(\mathbf{r}-\mathbf{R'}_{i})\approx\\
&\approx& \delta_{i,i'}\delta_{\alpha\alpha'}\mathbf{R}_{i}+\delta_{i,i'}\mathbf{d}_{\alpha\alpha'}.
\end{eqnarray} 
When $\mathbf{E}$ is applied in-plane:
\begin{eqnarray}
H^{\uparrow}_{d}=-q E_{a}(t)\sum_{i,\alpha\alpha'}\left(R^{a}_{i}\delta_{\alpha\alpha'}+d^{a}_{\alpha\alpha'}\right)c^{+}_{i,\alpha}c_{i,\alpha'}.\label{dipole2}
\end{eqnarray}
The symbols $(\uparrow, \downarrow)$ refer to the two $m_{J}$ projections in the model (\ref{latham}) and (\ref{SOham}), remembering that due to the presence of the spin-orbit coupling, the spin operator is not a properly well defined conserved quantity. In the more realistic case, $(\uparrow, \downarrow)$ will refer to the to the signs $(+,-)$ of the total angular momentum projections of the states from which the $|\uparrow\rangle$ and $|\downarrow\rangle$ states are built \cite{MR11}. The dipole matrix elements are defined as $d^{j}_{\alpha\alpha'}=\int d^{2}\mathbf{r}\phi^{*}_{\alpha}r^{j}\phi_{\alpha'}$. In our specific case, for the spin-up projection 
\begin{equation}
d^{j}_{\alpha\alpha'}=\int d^{2}\mathbf{x}\langle s| x^{j}|p_{-1}\rangle= d\left(\delta^{jx}+i \delta^{jy}\right),\label{dipole}
\end{equation}
where $d=\int d^{2}\mathbf{x}\langle s| x|p_{-1}\rangle$. The electric dipole transitions behind this result occur when $\Delta m_{J}=\pm 1$ and $\Delta J=\pm 1$\cite{G05}.

The first term in (\ref{dipole2}) is of the form $-q\mathbf{E}(t)\cdot\mathbf{R}_{i}$. The most direct way of dealing with this term is to change the gauge to the temporal gauge $A_{0}=0$ with the following phase change in the fermions: $c_{i}\rightarrow e^{i\Lambda_{i}(t)}c_{i}$, $c^{+}_{i}\rightarrow c^{+}_{i}e^{-i\Lambda_{i}(t)}$, with $\Lambda_{i}(t)=q\int_{t}\mathbf{E}(\tau)\mathbf{R}_{i}d\tau+\Lambda^{0}(\mathbf{R}_{i})\equiv q\mathbf{A}(t)\mathbf{R}_{i}+\Lambda^{0}(\mathbf{R}_{i})$. In absence of an external magnetic field, we have the freedom to choose $\Lambda^{0}(\mathbf{R}_{i})=0$ so the Hamiltonian (\ref{latham}), after reducing to a two component system, is now diagonal in momentum space: 
\begin{eqnarray}
\nonumber H_{\uparrow}(\mathbf{k},\mathbf{A})&=&d_{1}(\mathbf{k},\mathbf{A})\tau_{1}-d_{2}(\mathbf{k},\mathbf{A})\tau_{2}+\\
&+&\mu(\mathbf{k},\mathbf{A})\tau_{0}+ m(\mathbf{k},\mathbf{A})\tau_{3},\label{lathabmatrix}
\end{eqnarray}
with the parameters $m$, $\mu$, and $d_{i}$ defined in table \ref{tab:delements}.
\begin{table}
\caption{\label{tab:delements}Parameters for the BHZ model $\left(\mathbf{X}=a\mathbf{k}-qa\mathbf{A}\right)$}
\begin{ruledtabular}
\begin{tabular}{c}
$\mu(\mathbf{X})=\frac{1}{2}(\epsilon_{s}+\epsilon_{p}-\frac{\lambda}{2}-(2t_{s}+t_{p})\sum_{j=x,y}\cos(X_{j}))$ \\
$m(\mathbf{X})=\frac{1}{2}(\epsilon_{s}-\epsilon_{p}+\frac{\lambda}{2}-(2t_{s}-t_{p})\sum_{j=x,y}\cos(X_{j}))$ \\
$d_{1}(\mathbf{X})=\sqrt{2}t_{sp}\sin(X_{x})$\\
$d_{2}(\mathbf{X})=\sqrt{2}t_{sp}\sin(X_{y})$
\end{tabular}
\end{ruledtabular}
\end{table}

With this change of gauge and in matrix notation $H_{d}$ reads
\begin{eqnarray}
H^{\uparrow}_{d}=-q\frac{d}{c}(\dot{A}_{x}\tau_{1}-\dot{A}_{y}\tau_{2}).\label{dipole3}
\end{eqnarray}
Note that now the electronic states are coupled to the gauge vector field $\mathbf{A}$ trough the standard Peierls substitution in (\ref{lathabmatrix}) and trough a nonminimal coupling to the time derivative of  the vector field in (\ref{dipole3}). Indeed the Hamiltonian (\ref{dipole3}) is nothing but a way of writing the Hamiltonian of the Stark effect. 
In order to ease the computations, we will work in the continuum limit expanding (\ref{lathabmatrix}) around the $\Gamma$ point $\mathbf{k}=0$, and in the linear response regime corresponding to keep terms linear in $\mathbf{A}$ in (\ref{lathabmatrix}) and $\dot{\mathbf{A}}$ in (\ref{dipole3}). The Hamiltonian takes the form of a single specie of Dirac fermion (dropping an irrelevant redefinition of the zero of energies):
\begin{eqnarray}
H_{\uparrow}(\mathbf{k})=vk_{x}\tau_{1}-vk_{y}\tau_{2}+m\tau_{3},\label{effham0}
\end{eqnarray}
with $m=(\epsilon_{s}-\epsilon_{p}+\lambda/2+t_{p}-2t_{s})/2$ and $v=\sqrt{2}at_{sp}$. The total effective Hamiltonian describing the interaction between the electrons and the electromagnetic field is
\begin{eqnarray}
H^{\uparrow}_{A}=q \frac{v}{c}(A_{x}+\frac{d}{v}\dot{A}_{x})\tau_{1}-q \frac{v}{c}(A_{y}+\frac{d}{v}\dot{A}_{y})\tau_{2}.\label{effhamA}
\end{eqnarray}
A comment is in order here. Because we have fixed the gauge to the temporal gauge $A_{0}=0$ the system described by (\ref{effham0}) and (\ref{effhamA}) is not invariant under the entire gauge group, but still invariant under time independent gauge transformations. This is a standard situation when the Hamiltonian version of lattice gauge theories is considered\cite{KS75} and it will not be hard to find gauge invariant expressions when calculating the effective electromagnetic action.
If we define now the fermion field $\psi(x)=(s(x),p_{-1}(x))^{T}$ and the adjoint field as $\bar{\psi}=(s^{+}(x),p^{+}_{-1}(x))\gamma^{0}$, with $\gamma^{0}=\tau_{3}$, the effective low energy fermionic action reads:
\begin{eqnarray}
\nonumber\mathcal{S}_{f}&=&\int d^{3}x -i\bar{\psi}\gamma^{\mu}\partial_{\mu}\psi- m\bar{\psi}\psi-\\
&-&q\bar{\psi}\gamma^{\mu}\psi(A_{\mu}+\zeta^{\nu}F_{\nu\mu}).\label{effaction}
\end{eqnarray}
Above we have defined $\gamma^{1}=\tau_{3}\tau_{1}=i\tau_{2}$, and $\gamma^{2}=-\tau_{3}\tau_{2}=i\tau_{1}$. We have written an entirely gauge invariant fermionic action by defining a constitutive constant vector $\zeta^{\nu}=(d/v,0,0)$ in our specific problem. Note that the last component in the action (\ref{effaction}) has the form of a non minimal coupling between the fermionic current and the gauge field. Nonminimal couplings to the electromagnetic field are not so rare in Condensed Matter Physics. If an external magnetic field is applied to a spinful system, a Zeeman term of the form $H_{Z}=g\mu_{B}B_{a}\sum_{\mathbf{k},\alpha\alpha'}c^{+}_{\mathbf{k}\alpha}s^{a}_{\alpha\alpha'}c_{\mathbf{k},\alpha'}$ with $B_{a}=\epsilon_{abc}\partial_{b}A_{c}$,  would be needed to add to the Hamiltonian, or a similar term to (\ref{dipole3}) but coupling states with $\Delta m_{J}=0$, $\Delta J=\pm 1$ would appear if an electric field is applied perpendicularly to the sample leading, together with the SO coupling, to the Rashba term in the tight binding Hamiltonian\cite{KM05,TMS13}. The crucial difference with other non minimal couplings is that the fourth term in (\ref{effaction}) is generated by the external electric field and it will induce extra terms in the linear response regime.

\section{Modified QAH response}

Let us calculate the induced electronic current by computing the odd part of the effective field theory for the electromagnetic field. The effective action takes the form of a Chern-Simons action modified by the non minimal term\cite{R84}
\begin{eqnarray}
\nonumber \Gamma_{eff}&=&\int d^{3}x\sigma_{xy}\epsilon^{\mu\rho\nu}(A_{\mu}+\zeta^{\lambda}F_{\lambda\mu})\partial_{\rho}
(A_{\nu}+\zeta^{\sigma}F_{\sigma\nu})-\\
&-&J^{\mu}_{e}A_{\mu}.\label{effactionA}
\end{eqnarray}
In the continuum model $\sigma_{xy}=\frac{q^{2}}{4\pi}sign(m)$, while if we were used the full lattice model $\sigma_{xy}=\frac{q^{2}}{2\pi}sign(m)$. From the effective action (\ref{effactionA}) we can easily read out the induced electronic current:
\begin{eqnarray}
\langle J^{\mu}_{e}\rangle\equiv\frac{\delta\Gamma_{eff}}{\delta A_{\mu}}=\sigma_{xy}\epsilon^{\mu\rho\nu}\partial_{\rho}A_{\nu}+\sigma_{xy}\epsilon^{\mu\rho\nu}\zeta^{\sigma}\partial_{\rho}F_{\sigma\nu}.\label{inducedcurrent}
\end{eqnarray}
Note that the induced current (\ref{inducedcurrent}) is gauge invariant. The significance of the second term in the induced charge current (\ref{inducedcurrent}) is most apparent if we write it in components ($\mathbf{E}(t)$ pointing along the $y$ direction):
\begin{equation}
\langle J^{x}_{e}(t)\rangle=\sigma_{xy}E_{y}(t)+\sigma_{xy}\zeta^{0}\dot{E}_{y}(t).\label{currentcomponents}
\end{equation}
The Hall charge response of the system is now not just a term proportional to the applied electric field but it takes an extra contribution proportional to the time derivative of the electric field. This second term, although coming from the odd part of the polarization tensor and being proportional to $\sigma_{xy}$, is not universal because it is proportional to $\zeta^{0}$. This parameter with units of time turns out to be odd under time reversal symmetry. Another important observation is that if the electric field $\mathbf{E}$ is time independent, this second term is zero, so it will only be observable under  AC fields. It is important to stress that the new terms in (\ref{currentcomponents}) do not renormalize the odd part of the polarization tensor  which is the origin of (\ref{effactionA}). 

\section{Modified QSH response}

Now it is time to see how the previous results get modified when we include time reversal symmetry in the system by adding the spin other angular momentum projection in the BHZ model. For this projection the states to be considered are $s_{i}$ and $p_{i,1}$. In the basis formed by these two states the low energy momentum space Hamiltonian reads:
\begin{equation}
H_{\downarrow}(\mathbf{k})=-vk_{x}\tau_{1}-vk_{y}\tau_{2}+m\tau_{3}\equiv H^{*}_{\uparrow}(-\mathbf{k}),\label{spindownham}
\end{equation}
where $H_{\uparrow}(\mathbf{k})$ is defined in (\ref{effham0}). Following the same steps leading to the Hamiltonian (\ref{dipole3}) it is not difficult to show that for this projection, the dipole interaction takes the form
\begin{equation}
H^{\downarrow}_{d}=-q\frac{d}{c}(\dot{A}_{x}\tau_{1}+\dot{A}_{y}\tau_{2}),\label{dipoledown}
\end{equation}
which is nothing but the Hamiltonian (\ref{dipole3}) after applying the operation of time reversal inversion: $H^{\downarrow}_{d}=\mathcal{T}H^{\uparrow}_{d}\mathcal{T}^{-1}$. Let us write down the total low energy Hamiltonian for both projections:
\begin{equation}
H(\mathbf{k})=v s_{3}\tau_{1}k_{x}-vs_{0}\tau_{2}k_{2}+s_{0}\tau_{3}m.\label{totalham0}
\end{equation}
The Pauli matrices $(s_{0},\mathbf{s})$ stand for the two projections of $m_{J}$. If we want to go to a Lagrangian description of the problem we define as before an adjoint spinor $\bar{\psi}=\psi^{+}\gamma^{0}$, with $\gamma^{0}=s_{0}\tau_{3}$, and a set of $\gamma$ matrices as $\gamma^{1}=i s_{3}\tau_{2}$, and $\gamma^{2}=i s_{0}\tau_{1}$. Because we have now more than a single specie of fermions, we can define $\gamma_{5}=-i\gamma^{0}\gamma^{1}\gamma^{2}=s_{3}\tau_{0}$. With this choice of matrices, the two $m_{J}$ projections correspond to the two chiralities in the effective Hamiltonian. It is illuminating to see how the Hamiltonians (\ref{dipole3}) and (\ref{dipoledown}) read in terms of this set of matrices:
\begin{equation}
H_{d}=-q\frac{d}{c}\gamma_{5}\gamma^{i}\dot{A}_{i}.\label{dipoletotal}
\end{equation}
It means that the electronic states couple both minimally and non minimally to the gauge field $A_{\mu}$ and the latter is a chiral coupling. The fermionic action (\ref{effaction}) is properly modified to take into account both chiral species and the chiral coupling with $F_{\mu\nu}$:
\begin{eqnarray}
\nonumber\mathcal{S}_{f}&=&\int d^{3}x -i\bar{\psi}\gamma^{\mu}\partial_{\mu}\psi- m\bar{\psi}\psi-\\
&-&q\bar{\psi}\gamma^{\mu}(A_{\mu}+\gamma_{5}\zeta^{\nu}F_{\nu\mu})\psi.\label{effaction2}
\end{eqnarray}
In this case we have a different situation than the one described above when a spin polarized model was considered. Now due to the requirement of being the theory time reversal invariant electrons of different spin couple to the electric field through the dipole term with opposite sign, so, as it happens in the standard QSHE a spin current will be generated:
\begin{eqnarray}
\langle J^{\mu}_{s}\rangle\equiv\langle J^{\mu}_{\uparrow}-J^{\mu}_{\downarrow}\rangle=2\sigma_{xy}\epsilon^{\mu\rho\nu}\partial_{\rho}A_{\nu},\label{spincurrent}
\end{eqnarray}
but now we will also get a nonzero charge response coming from the dipole term:
\begin{eqnarray}
\langle J^{\mu}_{e}\rangle\equiv\langle J^{\mu}_{\uparrow}+J^{\mu}_{\downarrow}\rangle=2\sigma_{xy}\epsilon^{\mu\rho\nu}\zeta^{\sigma}\partial_{\rho}F_{\sigma\nu}.\label{inducedcurrent2}
\end{eqnarray}
The appearance of a charge response in the time reversal invariant system is not surprising actually. The important observation is that the two species of fermions couple oppositely to the electric field through the dipole term. This sign difference conspires with the opposite sign of the Berry phase in both species leading to a non vanishing value of $\langle J^{\mu}_{e}\rangle$. In this case, the role of the chiral field is played by $V_{\nu}=\zeta^{\sigma}F_{\sigma\nu}$\cite{CGV10,VAA13}. The expressions (\ref{currentcomponents}) and (\ref{spincurrent}-\ref{inducedcurrent2}) are the most important results of this work as long as they are the fingerprint of the anomalous modifications in the QAH and QSH phases. However, currently the expressions (\ref{currentcomponents}) and (\ref{spincurrent}-\ref{inducedcurrent2}) are not directly measurable. The experimental tests pass through performing transport measurements using the boundary states.

\section{Modified edge states}

Following the Callan-Harvey effect\cite{CH85}, let us consider a finite system $\Omega$ with a boundary $\partial\Omega$ described by the action (\ref{effactionA}) and apply a gauge transformation on the electromagnetic field $A_{\mu}\rightarrow A_{\mu}+\partial_{\mu}\Lambda(x)$. Note that under gauge transformations $F_{\lambda\sigma}$ remains invariant. The variation of the effective action (\ref{effactionA}) under the previous transformation can be written as
\begin{eqnarray}
\delta_{\Lambda}\Gamma_{eff}=\sigma_{xy}\int_{\partial\Omega}d^{2}\Sigma_{\mu}\Lambda(x)\epsilon^{\mu\rho\nu}\partial_{\rho}(A_{\nu}+\zeta^{\sigma}F_{\sigma\nu}),\label{gaugevariation}
\end{eqnarray}
where $d^{2}\Sigma_{\mu}$ stands for the differential surface element pointing perpendicular to the surface defined by $\partial\Omega$. Without loss of generality we can choose such surface to be defined by the coordinates $(t,x)$, so $d^{2}\Sigma_{\mu}$ will point along the $y$ direction. As usual, a non vanishing variation (\ref{gaugevariation}) means the system is not gauge invariant when confining it in a finite geometry. This lack of gauge invariance must be compensated by  another element in the system. In the standard case of a pure CS term, an one dimensional chiral massless fermion will appear at the boundary to restore gauge invariance (the so called chiral massless Schwinger model). Let us see how this chiral fermion changes to deal with the second term in (\ref{gaugevariation}). Let us write down a modified version of the fermionic action for the chiral Schwinger model\cite{H73,JR85} :
\begin{eqnarray}
\nonumber \mathcal{S}_{1+1}&=&\int d^{2}x -i\bar{\psi}\hat{\gamma}^{\mu}\partial_{\mu}\psi-\\
&-&q\bar{\psi}\hat{\gamma}^{\mu}P_{L}\psi(A_{\mu}+\zeta^{\nu}F_{\nu\mu}),\label{actionCSW}
\end{eqnarray}
with $\hat{\gamma}^{0}=\tau_{3}$, $\hat{\gamma}^{1}=i\tau_{1}$, $\hat{\gamma}_{5}=\tau_{2}$ and $P_{L}=\frac{1}{2}(1+\hat{\gamma}_{5})$. The presence of the projector $P_{L}$ means that only the left handed fermionic mode is coupled to $A_{\mu}$ and $\zeta^{\nu}F_{\nu\mu}$. Actually, the coupling between $A_{\mu}$ and the right or left fermion depends on the sign of $m$. If we integrate out fermions and apply the previous gauge transformation, the variation is nonzero (meaning that the theory is not gauge invariant) and takes the form:
\begin{eqnarray}
\delta_{\Lambda}\Gamma_{1+1}=-\frac{q^{2}}{4\pi}\int d^{2}x \Lambda(x)\epsilon^{\alpha\beta}\partial_{\alpha}(A_{\beta}+\zeta^{\sigma}F_{\sigma\beta}).\label{gaugevariation2}
\end{eqnarray}
which exactly cancels (\ref{gaugevariation}). The conclusion is clear: the edge mode in our system consists in a chiral massless Dirac fermion coupled both minimally and non minimally to the external electromagnetic field. When time reversal symmetry is present the one dimensional metal is not described by the chiral Schwinger model (\ref{actionCSW}) but by the Schwinger model with a nonminimal chiral coupling term:
\begin{eqnarray}
\nonumber\mathcal{S}_{HL}&=&\int d^{2}x -i\bar{\psi}\hat{\gamma}^{\mu}\partial_{\mu}\psi-\\
&-& q\bar{\psi}\hat{\gamma}^{\mu}(A_{\mu}+\hat{\gamma}_{5}\zeta^{\nu}F_{\nu\mu})\psi.\label{actionHSM}
\end{eqnarray}
Recent measurements probe the spin polarization of currents in the QSHE\cite{BRB12,NSB13}. We suggest to use similar techniques but allowing for time varying voltages to explore the consequences of the modifications of (\ref{actionCSW}) and (\ref{actionHSM})\footnote{The effect of anomalous terms in the Hall response can be tested by measuring the AC conductance $G(\omega)$ of the edge states. The precise computation of $G(\omega)$ will presented in a separate publication.}. 

%A lower bound for the parameter $\zeta^{0}$ can be found by using hydrogenic wavefunctions for high principal quantum number and the real %value for $v$, corresponding to inverse frequencies of the order of $200$ meV\cite{BHZ06}.

\section{Summary}

In the present work we have found how the Hall responses of the QAHE and the QSHE are modified when we take into account the intra atomic dipole elements between the states close to the Fermi level. These modifications, being non universal, are proportional to the Hall conductance $\sigma_{xy}$. We have also found how the edge states acquire an extra non minimal coupling term with the electromagnetic field. This non minimal term is different from the QAHE and the QSHE. These results can be observed by electrical conductance and tunnelling conductance measurements. The results presented here pave the way for new search of phenomena in the physics of two dimensional topological insulators.

\section{Aknowledgements}

The author acknowledges discussions with M. Sturla about the subject and H. Ochoa and B. Amorim for discussions in the early stages of this project.  The author aknowledges the JAE-doc program and the Spanish MEC through Grants No. FIS2011-23713 and No. PIB2010BZ-00512 for financial support.

%\bibliography{dipolebiblio}

\begin{thebibliography}{32}
\expandafter\ifx\csname natexlab\endcsname\relax\def\natexlab#1{#1}\fi
\expandafter\ifx\csname bibnamefont\endcsname\relax
  \def\bibnamefont#1{#1}\fi
\expandafter\ifx\csname bibfnamefont\endcsname\relax
  \def\bibfnamefont#1{#1}\fi
\expandafter\ifx\csname citenamefont\endcsname\relax
  \def\citenamefont#1{#1}\fi
\expandafter\ifx\csname url\endcsname\relax
  \def\url#1{\texttt{#1}}\fi
\expandafter\ifx\csname urlprefix\endcsname\relax\def\urlprefix{URL }\fi
\providecommand{\bibinfo}[2]{#2}
\providecommand{\eprint}[2][]{\url{#2}}

\bibitem[{\citenamefont{Thouless et~al.}(1982)\citenamefont{Thouless, Kohmoto,
  Nightingale, and den Nijs}}]{TKN82}
\bibinfo{author}{\bibfnamefont{D.~J.} \bibnamefont{Thouless}},
  \bibinfo{author}{\bibfnamefont{M.}~\bibnamefont{Kohmoto}},
  \bibinfo{author}{\bibfnamefont{M.~P.} \bibnamefont{Nightingale}},
  \bibnamefont{and} \bibinfo{author}{\bibfnamefont{M.}~\bibnamefont{den Nijs}},
  \bibinfo{journal}{Phys. Rev. Lett.} \textbf{\bibinfo{volume}{49}},
  \bibinfo{pages}{405} (\bibinfo{year}{1982}).

\bibitem[{\citenamefont{Kane and Mele}(2005)}]{KM05}
\bibinfo{author}{\bibfnamefont{C.~L.} \bibnamefont{Kane}} \bibnamefont{and}
  \bibinfo{author}{\bibfnamefont{E.~J.} \bibnamefont{Mele}},
  \bibinfo{journal}{Phys. Rev. Lett.} \textbf{\bibinfo{volume}{95}},
  \bibinfo{pages}{226801} (\bibinfo{year}{2005}).

\bibitem[{\citenamefont{Sinova et~al.}(2004)\citenamefont{Sinova, Culcer, Niu,
  Sinitsyn, Jungwirth, and MacDonald}}]{SCN04}
\bibinfo{author}{\bibfnamefont{J.}~\bibnamefont{Sinova}},
  \bibinfo{author}{\bibfnamefont{D.}~\bibnamefont{Culcer}},
  \bibinfo{author}{\bibfnamefont{Q.}~\bibnamefont{Niu}},
  \bibinfo{author}{\bibfnamefont{N.~A.} \bibnamefont{Sinitsyn}},
  \bibinfo{author}{\bibfnamefont{T.}~\bibnamefont{Jungwirth}},
  \bibnamefont{and} \bibinfo{author}{\bibfnamefont{A.~H.}
  \bibnamefont{MacDonald}}, \bibinfo{journal}{Phys. Rev. Lett.}
  \textbf{\bibinfo{volume}{92}}, \bibinfo{pages}{126603}
  (\bibinfo{year}{2004}).

\bibitem[{\citenamefont{Bernevig and Zhang}(2005)}]{BZ05}
\bibinfo{author}{\bibfnamefont{B.~A.} \bibnamefont{Bernevig}} \bibnamefont{and}
  \bibinfo{author}{\bibfnamefont{S.-C.} \bibnamefont{Zhang}},
  \bibinfo{journal}{Phys. Rev. B} \textbf{\bibinfo{volume}{72}},
  \bibinfo{pages}{115204} (\bibinfo{year}{2005}).

\bibitem[{\citenamefont{Haldane}(1988)}]{H88}
\bibinfo{author}{\bibfnamefont{F.~D.~M.} \bibnamefont{Haldane}},
  \bibinfo{journal}{Phys. Rev. Lett.} \textbf{\bibinfo{volume}{61}},
  \bibinfo{pages}{2015} (\bibinfo{year}{1988}).

\bibitem[{\citenamefont{Klitzing et~al.}(1980)\citenamefont{Klitzing, Dorda,
  and Pepper}}]{KDP80}
\bibinfo{author}{\bibfnamefont{K.~v.} \bibnamefont{Klitzing}},
  \bibinfo{author}{\bibfnamefont{G.}~\bibnamefont{Dorda}}, \bibnamefont{and}
  \bibinfo{author}{\bibfnamefont{M.}~\bibnamefont{Pepper}},
  \bibinfo{journal}{Phys. Rev. Lett.} \textbf{\bibinfo{volume}{45}},
  \bibinfo{pages}{494} (\bibinfo{year}{1980}).

\bibitem[{\citenamefont{König et~al.}(2007)\citenamefont{König, Wiedmann,
  Brüne, Roth, Buhmann, Molenkamp, Qi, and Zhang}}]{KWS07}
\bibinfo{author}{\bibfnamefont{M.}~\bibnamefont{König}},
  \bibinfo{author}{\bibfnamefont{S.}~\bibnamefont{Wiedmann}},
  \bibinfo{author}{\bibfnamefont{C.}~\bibnamefont{Brüne}},
  \bibinfo{author}{\bibfnamefont{A.}~\bibnamefont{Roth}},
  \bibinfo{author}{\bibfnamefont{H.}~\bibnamefont{Buhmann}},
  \bibinfo{author}{\bibfnamefont{L.~W.} \bibnamefont{Molenkamp}},
  \bibinfo{author}{\bibfnamefont{X.-L.} \bibnamefont{Qi}}, \bibnamefont{and}
  \bibinfo{author}{\bibfnamefont{S.-C.} \bibnamefont{Zhang}},
  \bibinfo{journal}{Science} \textbf{\bibinfo{volume}{318}},
  \bibinfo{pages}{766} (\bibinfo{year}{2007}).

\bibitem[{\citenamefont{Chang et~al.}(2013)\citenamefont{Chang, Zhang, Feng,
  Shen, Zhang, Guo, Li, Ou, Wei, Wang et~al.}}]{CZF13}
\bibinfo{author}{\bibfnamefont{C.-Z.} \bibnamefont{Chang}},
  \bibinfo{author}{\bibfnamefont{J.}~\bibnamefont{Zhang}},
  \bibinfo{author}{\bibfnamefont{X.}~\bibnamefont{Feng}},
  \bibinfo{author}{\bibfnamefont{J.}~\bibnamefont{Shen}},
  \bibinfo{author}{\bibfnamefont{Z.}~\bibnamefont{Zhang}},
  \bibinfo{author}{\bibfnamefont{M.}~\bibnamefont{Guo}},
  \bibinfo{author}{\bibfnamefont{K.}~\bibnamefont{Li}},
  \bibinfo{author}{\bibfnamefont{Y.}~\bibnamefont{Ou}},
  \bibinfo{author}{\bibfnamefont{P.}~\bibnamefont{Wei}},
  \bibinfo{author}{\bibfnamefont{L.-L.} \bibnamefont{Wang}},
  \bibnamefont{et~al.}, \bibinfo{journal}{Science}
  \textbf{\bibinfo{volume}{340}}, \bibinfo{pages}{167} (\bibinfo{year}{2013}).

\bibitem[{\citenamefont{Tsui et~al.}(1982)\citenamefont{Tsui, Stormer, and
  Gossard}}]{TSG82}
\bibinfo{author}{\bibfnamefont{D.~C.} \bibnamefont{Tsui}},
  \bibinfo{author}{\bibfnamefont{H.~L.} \bibnamefont{Stormer}},
  \bibnamefont{and} \bibinfo{author}{\bibfnamefont{A.~C.}
  \bibnamefont{Gossard}}, \bibinfo{journal}{Phys. Rev. Lett.}
  \textbf{\bibinfo{volume}{48}}, \bibinfo{pages}{1559} (\bibinfo{year}{1982}).

\bibitem[{\citenamefont{Grushin and de~Juan}(2012)}]{GJ12}
\bibinfo{author}{\bibfnamefont{A.~G.} \bibnamefont{Grushin}} \bibnamefont{and}
  \bibinfo{author}{\bibfnamefont{F.}~\bibnamefont{de~Juan}},
  \bibinfo{journal}{Phys. Rev. B} \textbf{\bibinfo{volume}{86}},
  \bibinfo{pages}{075126} (\bibinfo{year}{2012}).

\bibitem[{\citenamefont{Vald\'es~Aguilar
  et~al.}(2012)\citenamefont{Vald\'es~Aguilar, Stier, Liu, Bilbro, George,
  Bansal, Wu, Cerne, Markelz, Oh et~al.}}]{VSL12}
\bibinfo{author}{\bibfnamefont{R.}~\bibnamefont{Vald\'es~Aguilar}},
  \bibinfo{author}{\bibfnamefont{A.~V.} \bibnamefont{Stier}},
  \bibinfo{author}{\bibfnamefont{W.}~\bibnamefont{Liu}},
  \bibinfo{author}{\bibfnamefont{L.~S.} \bibnamefont{Bilbro}},
  \bibinfo{author}{\bibfnamefont{D.~K.} \bibnamefont{George}},
  \bibinfo{author}{\bibfnamefont{N.}~\bibnamefont{Bansal}},
  \bibinfo{author}{\bibfnamefont{L.}~\bibnamefont{Wu}},
  \bibinfo{author}{\bibfnamefont{J.}~\bibnamefont{Cerne}},
  \bibinfo{author}{\bibfnamefont{A.~G.} \bibnamefont{Markelz}},
  \bibinfo{author}{\bibfnamefont{S.}~\bibnamefont{Oh}}, \bibnamefont{et~al.},
  \bibinfo{journal}{Phys. Rev. Lett.} \textbf{\bibinfo{volume}{108}},
  \bibinfo{pages}{087403} (\bibinfo{year}{2012}).

\bibitem[{\citenamefont{Shimano et~al.}(2013)\citenamefont{Shimano, Yumoto,
  Yoo, Matsunaga, Tanabe, Hibino, Morimoto, and Aoki}}]{SYY13}
\bibinfo{author}{\bibfnamefont{R.}~\bibnamefont{Shimano}},
  \bibinfo{author}{\bibfnamefont{G.}~\bibnamefont{Yumoto}},
  \bibinfo{author}{\bibfnamefont{J.~Y.} \bibnamefont{Yoo}},
  \bibinfo{author}{\bibfnamefont{R.}~\bibnamefont{Matsunaga}},
  \bibinfo{author}{\bibfnamefont{S.}~\bibnamefont{Tanabe}},
  \bibinfo{author}{\bibfnamefont{H.}~\bibnamefont{Hibino}},
  \bibinfo{author}{\bibfnamefont{T.}~\bibnamefont{Morimoto}}, \bibnamefont{and}
  \bibinfo{author}{\bibfnamefont{H.}~\bibnamefont{Aoki}},
  \bibinfo{journal}{Nat. Commun.} \textbf{\bibinfo{volume}{4}},
  \bibinfo{pages}{1841} (\bibinfo{year}{2013}).

\bibitem[{\citenamefont{Chang}(2003)}]{C03}
\bibinfo{author}{\bibfnamefont{A.~M.} \bibnamefont{Chang}},
  \bibinfo{journal}{Rev. Mod. Phys.} \textbf{\bibinfo{volume}{75}},
  \bibinfo{pages}{1449} (\bibinfo{year}{2003}).

\bibitem[{\citenamefont{Grushin and Cortijo}(2011)}]{GC11}
\bibinfo{author}{\bibfnamefont{A.~G.} \bibnamefont{Grushin}} \bibnamefont{and}
  \bibinfo{author}{\bibfnamefont{A.}~\bibnamefont{Cortijo}},
  \bibinfo{journal}{Phys. Rev. Lett.} \textbf{\bibinfo{volume}{106}},
  \bibinfo{pages}{020403} (\bibinfo{year}{2011}).

\bibitem[{\citenamefont{Voigt}(1901)}]{V01}
\bibinfo{author}{\bibfnamefont{W.}~\bibnamefont{Voigt}},
  \bibinfo{journal}{Annalen der Physik} \textbf{\bibinfo{volume}{4}},
  \bibinfo{pages}{197} (\bibinfo{year}{1901}).

\bibitem[{\citenamefont{Nagaosa et~al.}(2010)\citenamefont{Nagaosa, Sinova,
  Onoda, MacDonald, and Ong}}]{NSO10}
\bibinfo{author}{\bibfnamefont{N.}~\bibnamefont{Nagaosa}},
  \bibinfo{author}{\bibfnamefont{J.}~\bibnamefont{Sinova}},
  \bibinfo{author}{\bibfnamefont{S.}~\bibnamefont{Onoda}},
  \bibinfo{author}{\bibfnamefont{A.~H.} \bibnamefont{MacDonald}},
  \bibnamefont{and} \bibinfo{author}{\bibfnamefont{N.~P.} \bibnamefont{Ong}},
  \bibinfo{journal}{Rev. Mod. Phys.} \textbf{\bibinfo{volume}{82}},
  \bibinfo{pages}{1539} (\bibinfo{year}{2010}).

\bibitem[{\citenamefont{Bernevig et~al.}(2006)\citenamefont{Bernevig, Hughes,
  and Zhang}}]{BHZ06}
\bibinfo{author}{\bibfnamefont{B.~A.} \bibnamefont{Bernevig}},
  \bibinfo{author}{\bibfnamefont{T.~L.} \bibnamefont{Hughes}},
  \bibnamefont{and} \bibinfo{author}{\bibfnamefont{S.-C.} \bibnamefont{Zhang}},
  \bibinfo{journal}{Science} \textbf{\bibinfo{volume}{314}},
  \bibinfo{pages}{1757} (\bibinfo{year}{2006}).

\bibitem[{\citenamefont{Fu and Kane}(2007)}]{FK07}
\bibinfo{author}{\bibfnamefont{L.}~\bibnamefont{Fu}} \bibnamefont{and}
  \bibinfo{author}{\bibfnamefont{C.~L.} \bibnamefont{Kane}},
  \bibinfo{journal}{Phys. Rev. B} \textbf{\bibinfo{volume}{76}},
  \bibinfo{pages}{045302} (\bibinfo{year}{2007}).

\bibitem[{\citenamefont{Liu et~al.}(2008)\citenamefont{Liu, Qi, Dai, Fang, and
  Zhang}}]{LQD08}
\bibinfo{author}{\bibfnamefont{C.-X.} \bibnamefont{Liu}},
  \bibinfo{author}{\bibfnamefont{X.-L.} \bibnamefont{Qi}},
  \bibinfo{author}{\bibfnamefont{X.}~\bibnamefont{Dai}},
  \bibinfo{author}{\bibfnamefont{Z.}~\bibnamefont{Fang}}, \bibnamefont{and}
  \bibinfo{author}{\bibfnamefont{S.-C.} \bibnamefont{Zhang}},
  \bibinfo{journal}{Phys. Rev. Lett.} \textbf{\bibinfo{volume}{101}},
  \bibinfo{pages}{146802} (\bibinfo{year}{2008}).

\bibitem[{\citenamefont{Yu et~al.}(2010)\citenamefont{Yu, Zhang, Zhang, Zhang,
  Dai, and Fang}}]{YZZ10}
\bibinfo{author}{\bibfnamefont{R.}~\bibnamefont{Yu}},
  \bibinfo{author}{\bibfnamefont{W.}~\bibnamefont{Zhang}},
  \bibinfo{author}{\bibfnamefont{H.-J.} \bibnamefont{Zhang}},
  \bibinfo{author}{\bibfnamefont{S.-C.} \bibnamefont{Zhang}},
  \bibinfo{author}{\bibfnamefont{X.}~\bibnamefont{Dai}}, \bibnamefont{and}
  \bibinfo{author}{\bibfnamefont{Z.}~\bibnamefont{Fang}},
  \bibinfo{journal}{Science} \textbf{\bibinfo{volume}{329}},
  \bibinfo{pages}{61} (\bibinfo{year}{2010}).

\bibitem[{\citenamefont{Michetti and Recher}(2011)}]{MR11}
\bibinfo{author}{\bibfnamefont{P.}~\bibnamefont{Michetti}} \bibnamefont{and}
  \bibinfo{author}{\bibfnamefont{P.}~\bibnamefont{Recher}},
  \bibinfo{journal}{Phys. Rev. B} \textbf{\bibinfo{volume}{83}},
  \bibinfo{pages}{125420} (\bibinfo{year}{2011}).

\bibitem[{\citenamefont{Griffiths}(2005)}]{G05}
\bibinfo{author}{\bibfnamefont{D.~J.} \bibnamefont{Griffiths}},
  \emph{\bibinfo{title}{Introduction to Quantum Mechanics}}
  (\bibinfo{publisher}{Pearson Prentice-Hall}, \bibinfo{year}{2005}).

\bibitem[{\citenamefont{Kogut and Susskind}(1975)}]{KS75}
\bibinfo{author}{\bibfnamefont{J.}~\bibnamefont{Kogut}} \bibnamefont{and}
  \bibinfo{author}{\bibfnamefont{L.}~\bibnamefont{Susskind}},
  \bibinfo{journal}{Phys. Rev. D} \textbf{\bibinfo{volume}{11}},
  \bibinfo{pages}{395} (\bibinfo{year}{1975}).

\bibitem[{\citenamefont{Tahir et~al.}(2013)\citenamefont{Tahir, Manchon,
  Sabeeh, and Schwingenschlogl}}]{TMS13}
\bibinfo{author}{\bibfnamefont{M.}~\bibnamefont{Tahir}},
  \bibinfo{author}{\bibfnamefont{A.}~\bibnamefont{Manchon}},
  \bibinfo{author}{\bibfnamefont{K.}~\bibnamefont{Sabeeh}}, \bibnamefont{and}
  \bibinfo{author}{\bibfnamefont{U.}~\bibnamefont{Schwingenschlogl}},
  \bibinfo{journal}{Applied Physics Letters} \textbf{\bibinfo{volume}{102}},
  \bibinfo{pages}{162412} (\bibinfo{year}{2013}).

\bibitem[{\citenamefont{Redlich}(1984)}]{R84}
\bibinfo{author}{\bibfnamefont{A.~N.} \bibnamefont{Redlich}},
  \bibinfo{journal}{Phys. Rev. Lett.} \textbf{\bibinfo{volume}{52}},
  \bibinfo{pages}{18} (\bibinfo{year}{1984}).

\bibitem[{\citenamefont{Cortijo et~al.}(2010)\citenamefont{Cortijo, Grushin,
  and Vozmediano}}]{CGV10}
\bibinfo{author}{\bibfnamefont{A.}~\bibnamefont{Cortijo}},
  \bibinfo{author}{\bibfnamefont{A.~G.} \bibnamefont{Grushin}},
  \bibnamefont{and} \bibinfo{author}{\bibfnamefont{M.~A.~H.}
  \bibnamefont{Vozmediano}}, \bibinfo{journal}{Phys. Rev. B}
  \textbf{\bibinfo{volume}{82}}, \bibinfo{pages}{195438}
  (\bibinfo{year}{2010}).

\bibitem[{\citenamefont{Vaezi et~al.}(2013)\citenamefont{Vaezi, Abedpour,
  Asgari, Cortijo, and Vozmediano}}]{VAA13}
\bibinfo{author}{\bibfnamefont{A.}~\bibnamefont{Vaezi}},
  \bibinfo{author}{\bibfnamefont{N.}~\bibnamefont{Abedpour}},
  \bibinfo{author}{\bibfnamefont{R.}~\bibnamefont{Asgari}},
  \bibinfo{author}{\bibfnamefont{A.}~\bibnamefont{Cortijo}}, \bibnamefont{and}
  \bibinfo{author}{\bibfnamefont{M.~A.~H.} \bibnamefont{Vozmediano}},
  \bibinfo{journal}{Phys. Rev. B} \textbf{\bibinfo{volume}{88}},
  \bibinfo{pages}{125406} (\bibinfo{year}{2013}).

\bibitem[{\citenamefont{Callan and Harvey}(1985)}]{CH85}
\bibinfo{author}{\bibfnamefont{C.~C.} \bibnamefont{Callan}} \bibnamefont{and}
  \bibinfo{author}{\bibfnamefont{J.}~\bibnamefont{Harvey}},
  \bibinfo{journal}{Nuclear Physics B} \textbf{\bibinfo{volume}{250}},
  \bibinfo{pages}{427 } (\bibinfo{year}{1985}).

\bibitem[{\citenamefont{Hagen}(1973)}]{H73}
\bibinfo{author}{\bibfnamefont{C.}~\bibnamefont{Hagen}},
  \bibinfo{journal}{Annals of Physics} \textbf{\bibinfo{volume}{81}},
  \bibinfo{pages}{67 } (\bibinfo{year}{1973}).

\bibitem[{\citenamefont{Jackiw and Rajaraman}(1985)}]{JR85}
\bibinfo{author}{\bibfnamefont{R.}~\bibnamefont{Jackiw}} \bibnamefont{and}
  \bibinfo{author}{\bibfnamefont{R.}~\bibnamefont{Rajaraman}},
  \bibinfo{journal}{Phys. Rev. Lett.} \textbf{\bibinfo{volume}{54}},
  \bibinfo{pages}{1219} (\bibinfo{year}{1985}).

\bibitem[{\citenamefont{Brune et~al.}(2012)\citenamefont{Brune, Roth, Buhmann,
  Hankiewicz, Molenkamp, Maciejko, Qi, and Zhang}}]{BRB12}
\bibinfo{author}{\bibfnamefont{C.}~\bibnamefont{Brune}},
  \bibinfo{author}{\bibfnamefont{A.}~\bibnamefont{Roth}},
  \bibinfo{author}{\bibfnamefont{H.}~\bibnamefont{Buhmann}},
  \bibinfo{author}{\bibfnamefont{E.~M.} \bibnamefont{Hankiewicz}},
  \bibinfo{author}{\bibfnamefont{L.~W.} \bibnamefont{Molenkamp}},
  \bibinfo{author}{\bibfnamefont{J.}~\bibnamefont{Maciejko}},
  \bibinfo{author}{\bibfnamefont{X.-L.} \bibnamefont{Qi}}, \bibnamefont{and}
  \bibinfo{author}{\bibfnamefont{S.-C.} \bibnamefont{Zhang}},
  \bibinfo{journal}{Nat. Phys.} \textbf{\bibinfo{volume}{8}},
  \bibinfo{pages}{485} (\bibinfo{year}{2012}).

\bibitem[{\citenamefont{Nowack et~al.}(2013)\citenamefont{Nowack, Spanton,
  Baenninger, Kronig, Kirtley, Kalisky, Ames, Leubner, Brune, Buhmann
  et~al.}}]{NSB13}
\bibinfo{author}{\bibfnamefont{K.~C.} \bibnamefont{Nowack}},
  \bibinfo{author}{\bibfnamefont{E.~M.} \bibnamefont{Spanton}},
  \bibinfo{author}{\bibfnamefont{M.}~\bibnamefont{Baenninger}},
  \bibinfo{author}{\bibfnamefont{M.}~\bibnamefont{Kronig}},
  \bibinfo{author}{\bibfnamefont{J.~R.} \bibnamefont{Kirtley}},
  \bibinfo{author}{\bibfnamefont{B.}~\bibnamefont{Kalisky}},
  \bibinfo{author}{\bibfnamefont{C.}~\bibnamefont{Ames}},
  \bibinfo{author}{\bibfnamefont{P.}~\bibnamefont{Leubner}},
  \bibinfo{author}{\bibfnamefont{C.}~\bibnamefont{Brune}},
  \bibinfo{author}{\bibfnamefont{H.}~\bibnamefont{Buhmann}},
  \bibnamefont{et~al.}, \bibinfo{journal}{Nat. Mater.}
  \textbf{\bibinfo{volume}{12}}, \bibinfo{pages}{787} (\bibinfo{year}{2013}).

\end{thebibliography}

\end{document}